\documentclass[5p]{elsarticle}
\usepackage{amsmath,amssymb}
\usepackage{graphicx}
\usepackage{xcolor}
\usepackage{times}

\newcommand{\sst}{\scriptscriptstyle}

\journal{Physics Letters B}

\begin{document}
\begin{frontmatter}
\title{Spectral-statistics properties of the experimental and theoretical light meson spectra}
\author[uno]{L. Mu\~noz}
\ead{laura@nuc5.fis.ucm.es}
\author[uno,dos]{C. Fern\'andez-Ram\'{\i}rez}
\ead{cesar@nuc2.fis.ucm.es}
\author[tres,cuatro]{A. Rela\~no}
\ead{armando.relano@gmail.com}
\author[dos]{J. Retamosa}
\address[uno]{European Centre for Theoretical Studies in Nuclear Physics and Related Areas (ECT*), 
Villa Tambosi, Strada delle Tabarelle 286, Villazzano (TN), I-38123, Italy}
\address[dos]{Grupo de F\'{\i}sica Nuclear, Departamento de F\'{\i}sica At\'omica, Molecular y Nuclear, 
Facultad de Ciencias F\'{\i}sicas, Universidad Complutense de Madrid, 
Avda. Complutense s/n, E-28040 Madrid, Spain}
\address[tres]{Instituto de Estructura de la Materia, IEM-CSIC, Serrano 123, E-28006 Madrid, Spain}
\address[cuatro]{Departamento de F\'{\i}sica Aplicada I,
Facultad de Ciencias F\'{\i}sicas, Universidad Complutense de Madrid, 
Avda. Complutense s/n, E-28040 Madrid, Spain}

\begin{abstract}

We present a robust analysis of the spectral fluctuations exhibited by the light meson spectrum.
This analysis provides information about the degree of chaos in light mesons
and may be useful to get some insight on the underlying interactions. Our analysis unveils that the
statistical properties of the light meson spectrum are close, but not exactly equal, to those of
chaotic systems. In addition, we have analyzed several theoretical spectra including the latest lattice QCD
calculation.  With a single exception, their statistical properties are close to those of a generic integrable
system, and thus incompatible with the experimental spectrum.

\end{abstract}

\end{frontmatter}

\section{Introduction} \label{sec:introduction}

Since Yukawa first proposed the pion to explain the internucleon force \cite{Yukawa35} and mesons
started to populate the Review of Particle Physics (RPP) \cite{PDG2010}, meson spectroscopy has
played a central role in the study of the strong interaction helping on its understanding and the
development of Quantum Chromodynamics (QCD).  Mesons constitute the simplest bound states for quarks
and gluons, and their accurate description is one of the principal aims of QCD. However, so far a
quantitative and predictive theory of confined states has not been achieved, hence, in order to
study the properties of mesons we have to rely on models which have to be consistent with the
underlying QCD.  Constituent quark models \cite{Godfrey98} are examples of this kind of modeling.

In the last decade an enormous experimental effort has been made in meson spectroscopy with several
facilities conducting research programs \cite{PhysRep} whose main goal has been to find exotic
mesons \cite{exotics} which do not fit within the quark-antiquark picture of quark models.  This
search has been fruitless so far but has put meson physics at the forefront of scientific research,
becoming a thriving research area with experimental collaborations in several facilities ---i.e. BES
(China) \cite{BES}, CLAS at JLab (USA) \cite{CLAS}, COMPASS at CERN (Switzerland) \cite{COMPASS},
J-PARC (Japan) \cite{Jparc} and Hall D under construction at JLab (USA) \cite{HallD}.

Theoretical research has not been oblivious to this experimental interest and several quark models
of mesons have made their appearance in the literature \cite{Koll00,Vijande05,Ebert09} trying to
match the low-lying experimental spectrum and complementing the classic calculation by Godfrey and
Isgur \cite{Godfrey85}.  Among the theoretical developments, it is noteworthy the lattice QCD
calculation of the meson spectrum by the Hadron Spectrum Collaboration (HSC) at JLab
\cite{Dudek,Dudek11}, although with the drawback of being computed at a high pion mass of 396 MeV.

Mesons can be considered as aggregates of quarks and
gluons. Therefore, the mass spectrum of low-lying mesons can be
understood as the energy spectrum of a quantum system, like
an atomic nucleus, and it consists on all the possible states which
stem from an interacting quantum system. Since Wigner
discovered that the statistical properties of complex nuclear spectra
are well described by the Gaussian Orthogonal Ensemble (GOE) of 
Random Matrix Theory \cite{Porter},
statistical methods have become a powerful tool to study the energy
spectra of quantum systems \cite{Gomez11,Guhr98}.  The most striking
result in this field is that the statistical properties of the
energy-level fluctuations determine if a system is chaotic, integrable
or intermediate. Moreover, the energy-level fluctuations of integrable
and chaotic systems are universal. The former display a non-correlated
sequence of levels, which follows the Poisson distribution
\cite{Berry77}, whereas the latter are characterized by a correlation
structure described by GOE \cite{Bohigas84}.  
This kind of analysis
has been already applied to the hadron mass spectrum in
\cite{Pascalutsa03} obtaining a chaotic-like behavior.  In
\cite{Fernandez07} the spectral-statistic techniques have been used to
compare the experimental baryon spectrum with theoretical ones,
focusing on the problem of missing resonances. The main conclusion of
this work is that quark models give rise to integrable spectra, whereas
the experimental one is chaotic. This result is not compatible with the
existence of missing resonances, predicted by the models and not observed
in the experiments. On the contrary, the lack of chaos in the models
shows that some ingredients are missing in them, as they are unable to
reproduce the experimental spectrum. As a corollary of this
work, it is stated that the theoretical models must reproduce the
spectral fluctuations of the experimental spectra, since they
determine the dynamical regime and the complexity of the real
interactions.

In this Letter we employ an improved version of the approach in
\cite{Fernandez07} to extend the work to mesons. We infer
the dynamical regime of light mesons from the statistical properties of their
spectrum. We also study a number of theoretical models based upon constituent
quarks \cite{Koll00,Vijande05,Ebert09,Godfrey85}, and the lattice QCD
calculation by the HSC at JLab \cite{Dudek,Dudek11}.

\section{Experimental spectrum}\label{sec:experiment}

\subsection{Statistical analysis and results}\label{ssec:results}

Prior to any statistical analysis of the spectral fluctuations one has to accomplish some
preliminary tasks. First of all, it is necessary to take into account all the symmetries that
properly characterize the system. Mixing different symmetries, i.e.  energy levels with different
values of the good quantum numbers, spoils the statistical properties deflecting them
  towards the Poisson statistics (see \cite{Porter,Guhr98} for generic reviews and \cite{Molina06}
for a recent work where the effects of both mixing symmetries and missing levels in the same
sequence are surveyed).  Hence, it is necessary to separate the whole spectrum into sequences of
energy levels involving the same symmetries (good quantum numbers). The usual symmetries associated
to mesons are spin ($J$), isospin ($I$), parity ($P$), $C-$parity ($C$) and strangeness. Strangeness
can be dropped due to the assumption of flavor $SU(3)$ invariance and because strange mesons
correspond to $I=1/2$ while the rest of the mesons considered in this Letter are $I=0,1$ (we do not
include in our analysis mesons with $c$, $b$ or $t$ quark content).  Therefore, the meson spectrum
is split into sequences with fixed values of $J$, $I$, $P$ and $C$.  In this work we take $d=126$
energy levels from the RPP experimental spectrum up to $2.5$ GeV of energy, which are distributed in
$n=23$ sequences of lengths $l_i \ge 3$, so that $\sum_{i=1}^n l_i =d$.

The energy spectrum of a quantum system is fully characterized by its level density $\rho(E)$. It
can be split into a smooth part $\overline{\rho}(E)$, giving the secular behavior with the energy,
and a fluctuating part $\widetilde{\rho}(E)$, which is responsible for the statistical properties of
the spectrum. The standard procedure to remove the smooth component of the level density is the
\textit{unfolding}.  Once this procedure has been performed the fluctuations from different systems
or from different parts of a single spectrum can be compared. Since the experimental meson spectrum
has been divided in very short sequences of levels, we have to use the so-called local unfolding
procedure \cite{Bae92} which considers that the variation of $\overline{\rho}(E)$ along a short
sequence of energy levels must be negligible, a reasonable assumption in the present case.  The
procedure is as follows.  Let $\{ E_i,\;i=1,2,\dots,l_x \}_X$ be an energy-level sequence
characterized by the set $X$ of good quantum numbers. Then, the distances between consecutive
levels, $S_i = E_{i+1} - E_i$, are rescaled using their average value $\left<S \right> =
\left(l_x-1\right)^{-1}\sum_{i=1}^{l_x-1} S_i$ to obtain the quantities $s_i =S_i / \left<S
\right>$, called generically nearest neighbor spacings (NNS). For the rescaled
  spectrum the mean level density $\overline{\rho}(E)= 1$, and therefore $\left<s
  \right>= \left(\overline{\rho}(E)\right)^{-1}=1$.

\begin{figure}
\begin{center}
\rotatebox{0}{\scalebox{0.5}[0.5]{\includegraphics{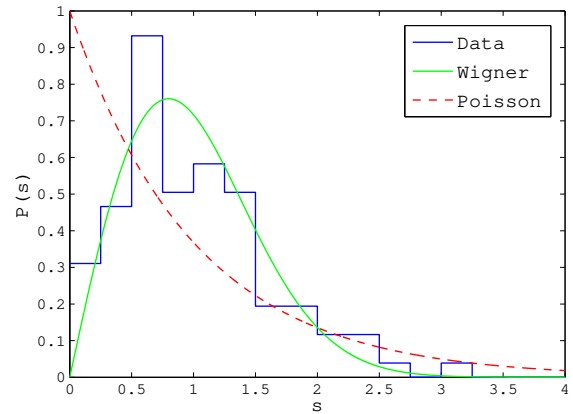}}}
\end{center}
\caption{(color online.) NNSD for the experimental spectrum (histogram) from the RPP, compared to the
  Wigner surmise (solid) and the Poisson distribution (dashed).}
\label{NNSDexp1}
\end{figure}

In this Letter, the statistical properties of the NNS are studied by means of the nearest neighbor
spacing distribution (NNSD) \cite{Mehta04}, denoted $P(s)$, which gives the probability that the
spacing between two consecutive unfolded levels lies between $s$ and $s+ds$. The NNSD follows the
Poisson distribution $P_p(s) = \exp(-s)$ for generic integrable systems \cite{Berry77} while chaotic
systems with time reversal and rotational invariance are well described by the GOE of random
matrices, whose NNSD follows the Wigner surmise $P_W(s)=\frac{\pi s}{2} \exp \left(-\frac{\pi
  s^2}{4} \right)$ \cite{Bohigas84}. Figure \ref{NNSDexp1} compares the $P(s)$ distribution for the
experimental spectrum with the Poisson distribution $P_P(s)$ and the Wigner surmise
$P_W(s)$. The experimental distribution is intermediate between the Poisson
  and Wigner limits, albeit it seems closer to the latter.

\begin{figure}
\begin{center}
\rotatebox{0}{\scalebox{0.5}[0.5]{\includegraphics{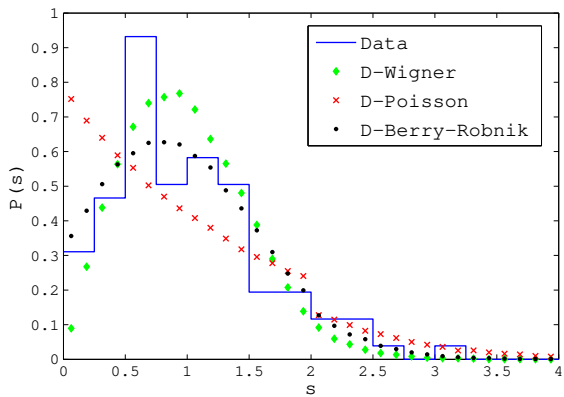}}}
\end{center}
\caption{(color online.) NNSD for the experimental spectrum (histogram) from the RPP, 
compared to the distorted distributions: Wigner (D-Wigner, diamonds), 
Poisson (D-Poisson, crosses) and Berry-Robnik with $f=0.78$ 
(D-Berry-Robnik, dots).}
\label{NNSDexp2}
\end{figure}

Before we proceed with a quantitative analysis, it is convenient to study the possible spurious
effects of the unfolding because the combination of the local unfolding with very short sequences of
levels may distort the actual $P(s)$ distribution \cite{AbulMagd06}.  Since $\left<s \right>=1$ for every spacing
sequence, no spacing can be greater than $l-1$, where $l$ is the sequence length, and therefore the
$P(s)$ distribution must exhibit a sharp cutoff at $s=l-1$.  When $l$ is large enough this cutoff is
irrelevant due to the exponential and Gaussian decays of the Poisson and Wigner
distributions. Obviously, this is not the case for smaller values of $l$; in order to take into
account this problem we will, as in \cite{Fernandez07}, \textquotedblleft distort\textquotedblright
\: Wigner and Poisson predictions including the effects of the unfolding procedure in the same way
as the experimental distribution. To generate
these distributions we divide two generic GOE and Poisson spectra of dimension $d=126$ in $23$ level
sequences with the same lengths of the experimental spectrum, and calculate their spacings. Then, we take a step further with respect to \cite{Fernandez07} and
instead of using just one spectrum in each case, we obtain the smooth behavior of the distorted distributions by averaging over $1000$ realizations
of the spectra. The corresponding distorted Poisson and Wigner NNSDs are denoted
by $P_{DP}(s)$ and $P_{DW}(s)$, and it is important to stress here that they will play from now on
the role of \textit{theoretical distributions} with which the data have to be compared. Figure \ref{NNSDexp2} displays the shape of these curves and a comparison with
figure \ref{NNSDexp1} shows that the unfolding effect is quite noticeable for $P_P(s)$.

Figure \ref{NNSDexp2} confirms that, after taking care of the unfolding distorsions, the statistical
properties of the experimental spectrum are intermediate between the Poisson and Wigner
predictions. In order to assess how close the fluctuations are to one limit or the other, some
distributions can be used to describe intermediate NNSD. One of the most frequently used is the
Berry-Robnik distribution $P_{BR}(f,s)$ \cite{BerryRobnik}, where $f$ represents the fraction of
phase-space volume dominated by chaotic orbits in the semiclassical limit. Of course, for a proper
comparison, we need the corresponding distorted Berry-Robnik distribution $P_{DBR}(f,s)$, which we
compute following the same procedure employed to obtain $P_{DP}(s)$ and $P_{DW}(s)$. Fitting the experimental $P(s)$ to $P_{DBR}(f,s)$ we obtain for the fractional density $f = 0.78 \pm 0.13$. Figure \ref{NNSDexp2} also displays $P_{DBR}(0.78,s)$.

Additional information can be gained by performing the K-S distribution test \cite{KS}. The
reference distribution is either $P_{DP}(s)$, $P_{DW}(s)$ or $P_{DBR}(0.78,s)$, and the null
hypothesis is that the experimental $P(s)$ distribution coincides with the one selected as
reference, against the hypothesis that both distributions are different.  The results for the
$p$-value obtained in each case are $p_{\sst DP} = 0.13$, $p_{\sst DW} = 0.47$ and $p_{\sst DBR} =
0.68$, respectively. This result is fully compatible with our previous analysis of the NNSD. The
statistical properties of the experimental meson spectrum are intermediate between the Poisson and
Wigner limits.  However, they are closer to the latter since a Berry-Robnik distribution with
$f=0.78$ fits well the experimental NNSD and $p_{\sst DBR} = 0.68$. It is also worth to note that
$p_{\sst DP}= 0.13$ is close to the usual limit for the null hypothesis to be rejected ($p \lesssim
0.10$).

\begin{figure}
\begin{center}
\rotatebox{0}{\scalebox{0.5}[0.5]{\includegraphics{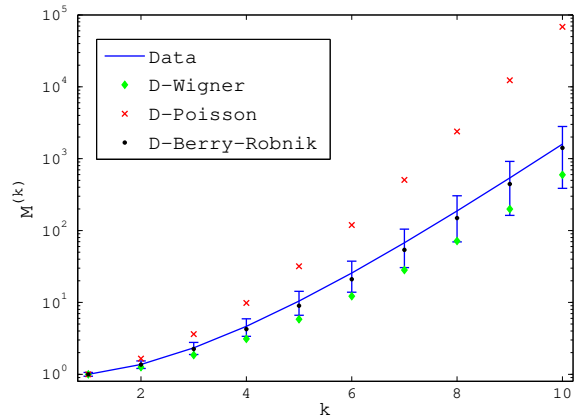}}}
\end{center}
\caption{(color online.) Moments of the NNSD, $M^{(k)}$, 
for the experimental spectrum (solid line with error bars), 
from the RPP, compared to those of the distorted distributions: 
Wigner (diamonds), Poisson (crosses)
and Berry-Robnik with $f=0.78$ (dots). }
\label{moments_exp}
\end{figure}

As a complementary test, we calculate the moments of the distributions. Gathering together all
the spacings $s_i$, the $k$-th moment, $M^{(k)}$ is calculated as $M^{(k)} = (d-n)^{-1}
\sum_{i=1}^{d-n} s_i^k$, where $d$ stands for the spectrum dimension and $n$ is the number of
spacing sequences.  Figure \ref{moments_exp} displays the moments $M^{(k)}\!, \; 1 \le k \le 10$ for
the experimental spacing distribution (the error bars correspond to the standard deviation), as well
as the $M^{(k)}$ corresponding to the distorted distributions $P_{DP}(s)$, $P_{DW}(s)$ and
$P_{DBR}(s)$. It is shown that the moments of the distorted Poisson distribution are outside and far
away from the error bars. Although the moments of $P_{DW}(s)$ are compatible with the experimental
data, they fall just on the lower edge of the error bars. Only the moments of $P_{DBR}(f,s)$ with
$f=0.78$ match the experimental result supporting our choice of $f$.

\begin{figure}
\begin{center}
\rotatebox{0}{\scalebox{0.5}[0.5]{\includegraphics{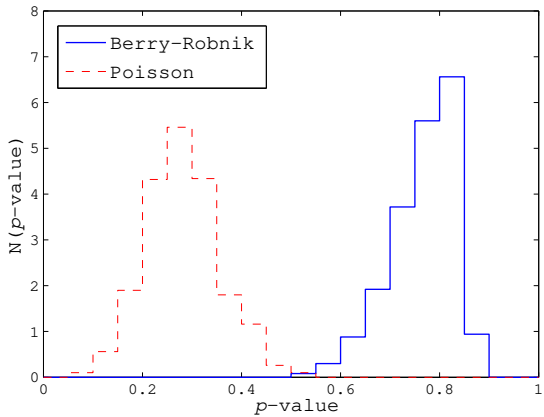}}}
\end{center}
\caption{(color online.) Distributions of the $p$-values of the K-S test for the 1000
\textquotedblleft realizations\textquotedblright \: of the
experimental spectrum within the error bars, for the null hypothesis that the distribution
coincides with Poisson (dashed histogram) and with Berry-Robnik for $f=0.78$ (solid histogram).}
\label{KSdistribution}
\end{figure}

Finally, it remains to test if our analysis is robust against the inclusion of the error bars
associated to the experimental data.  Considering the experimental meson energies as Gaussian random
variables with mean equal to the RPP estimation and variance equal to the error bar, we have
generated $1000$ \textquotedblleft realizations\textquotedblright \: of the experimental spectrum
and performed for each of them the K-S test.  If the energy levels are allowed to fluctuate
independently (in our case these fluctuations are induced by the error bars), the correlations are
usually weakened, displacing the statistics towards Poisson.  Therefore, it is appropriate to use as
reference distributions $P_{DP}(s)$ and $P_{DBR}(0.78,s)$.  Figure \ref{KSdistribution} shows that
the histograms of the resulting $p$-values are separated with almost no overlap. The distribution of
$p_{DBR}$-values is concentrated in the upper half with $\left<p_{\sst DBR}\right> = 0.77 \pm 0.07$,
and the histogram of the $p_{DP}$-values lies in the lower half with centroid $\left<p_{\sst
  DP}\right> = 0.28 \pm 0.07$.  It is important to notice that for almost every \textquotedblleft
realization\textquotedblright \: of the experimental spectrum $p_{\sst DBR} > p_{\sst DP}$,
sustaining the good agreement of the experiment with the Berry-Robnik distribution for $f=0.78$. For
the sake of completeness we have also used as reference distribution the Wigner surmise, obtaining
$\left<p_{\sst DW}\right> = 0.46 \pm 0.09$.

To summarize, our analysis is fairly robust against experimental errors and allows us to conclude
that the statistical properties of the experimental spectrum are intermediate between the Wigner and
Poisson limits, closer to the former and safely incompatible with the latter. That is, mesons are much closer to chaotic systems than to integrable ones. Moreover, a
Berry-Robnik distribution with 78\% of chaos provides the best description of the experimental
NNSD.

\subsection{Statistical significance of the analysis} \label{ssec:significance}

The statistical analysis performed in the previous section accounts for the
shortness of the sequences of levels through comparing to the theoretical
\textit{ad hoc} distorted NNS distributions. Also the size of the whole
spectrum is not too large and, though spectra of this kind can be found in
the literature (i.e. \cite{AbulMagd06} and references therein), a careful
analysis of the statistical significance is in order.

\begin{table}
\begin{center}
\caption[]{$p$-values of the K-S test for the experimental data from RPP and
the random ensemble of spectra with 78\% of chaos.
The null hypotheses are that the NNSD coincides with the distorted Wigner
surmise ($ p_{\sst DW}$), the distorted Poisson distribution ($ p_{\sst DP}$)
and the distorted Berry-Robnik with $f=0.78$ ($ p_{\sst DBR}$).} \vspace*{0.3cm}

\begin{tabular}{c||c|c}
\hline
     &          &      \\
     &  RPP  (data)   &  Random Ensemble  \\  \hline
     &          &      \\
$p_{\sst DW}$   & $0.47$ & $0.48 \pm 0.17$ \\ 
$p_{\sst DP}$   & $0.13$ & $0.25 \pm 0.13$ \\
$p_{\sst DBR}$  & $0.68$ & $0.71 \pm 0.09$ \\ \hline
\end{tabular}
\label{tableTest}
\end{center}
\end{table}

\begin{figure}
\begin{center}
\rotatebox{0}{\scalebox{0.5}[0.5]{\includegraphics{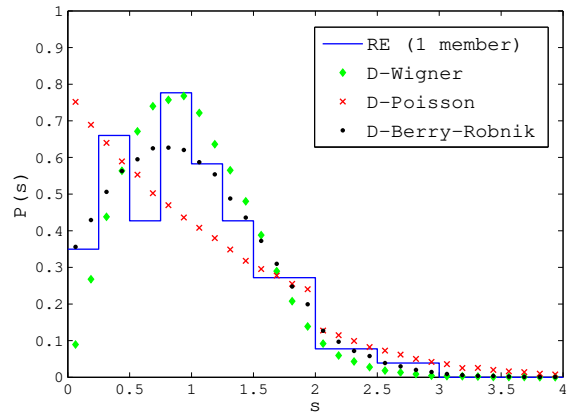}}}
\end{center}
\caption{(color online.) NNSD for one single realization of the RE (histogram) 
compared to the distorted distributions: Wigner (D-Wigner, diamonds), 
Poisson (D-Poisson, crosses) and Berry-Robnik with $f=0.78$ 
(D-Berry-Robnik, dots).}
\label{NNSD_RE}
\end{figure}

In the previous section we have found that the experimental spectrum is closer
to a chaotic system than to an integrable one, actually, obtaining the best
description employing a Berry-Robnik distribution with a 78\% of chaos. 
To test the validity of the analysis one should start from a spectrum with the
same size and structure (organized with the same number and length of the
sequences) of the experimental one, explicitly built with 78\% of chaos,
and perform the analysis in order to see how safely can we arrive
to the same conclusion. Then we build a random ensemble (RE) of 1000 
spectra with a 78\% of chaos, construct the NNSD in each case and run K-S tests
against the NNS distributions $P_{DW}(s)$, $P_{DP}(s)$ and $P_{DBR}(0.78,s)$.
We obtain the mean $p$-values and error bars (standard deviations)
shown in table \ref{tableTest} together with the ones obtained for the
actual experimental spectrum from the previous section.
In Fig. \ref{NNSD_RE} we show the NNSD for one of the spectra of the RE,
where it can be seen that is similar to the experimental one, that is,
the histogram is not very smooth due to the small size of the sample and
the $p$-values are: $p_{\sst DW} = 0.49$, $p_{\sst DP} = 0.20$
and $p_{\sst DBR} = 0.74$.

From the mean $p$-values one can conclude that a typical spectrum
of the ensemble is closer to chaotic systems than to integrable ones and is
best fitted with a mixed distribution with 78\% of chaos. From the error bars
we can see that the range of $p$-values for Wigner is the broadest but the
values inside the interval are such that the Wigner null hypothesis cannot be
rejected. The Poisson hypothesis, although the range of $p$-values is lower
(the lower edge is close to rejection), cannot be dismissed either. Thus,
most probably the actual distribution lies in between both,
and hence the case for trying a Berry-Robnik distribution, as we did for the
experimental one. When the K-S test is run for a Berry-Robnik with $f=0.78$,
the highest $p$-values of all three are obtained within a narrow range.
Then, we can deduce that in most of the cases we would arrive to the same
conclusion as for the experimental spectrum. But in order to better quantify
the statistical significance one should look at each individual realization.

It is important to compare the three $p$-values in each individual case
to see what would be the final decision in each one.
If we do this we find that in 80\% of the cases
$p_{\sst DW} > p_{\sst DP} $ and in 90\% $p_{\sst DBR}$ is the highest of
the three. The Poisson hypothesis can be rejected ($p < 0.1$) in 10\% of
the cases, while the Wigner in only a 0.5\% and the Berry-Robnik in none of
them. Hence, at a confidence level of approximately 90\%
we are recovering the correct result: the spectrum lies in between the Wigner
and Poisson extremes and the Berry-Robnik distribution with $f=0.78$ is the
best description of the spectrum. Additionally, we have run K-S tests for other
values of $f$ below and above $f=0.78$ checking that in all the cases lower
$p$-values are obtained. In summary, the methodology employed to
analyze the experimental spectrum is reliable and robust.

\section{Theoretical spectra}\label{sec:theory}

Next we analyze six theoretical calculations of the light meson spectrum and compare them to the
results from the previous section. These are: (i) The classic model by Godfrey and Isgur (set GI)
\cite{Godfrey85}, which is a relativized quark model where the interaction is built employing a one
gluon exchange potential and confinement is achieved through a spin-independent linear potential;
(ii) and (iii) are the fully relativistic quark models by Koll \textit{et al.} (sets K1 and K2 which
correspond, respectively, to models $\cal{A}$ and ${\cal B}$ in \cite{Koll00}) based on the
Bethe-Salpeter equation in its instantaneous approximation, a flavor-dependent two-body interaction
and a spin-dependent confinement force, being the latter the difference between the two models; (iv) the
relativistic quark model by Ebert \textit{et al.} (set E) \cite{Ebert09} based on a quasipotential
(this calculation has the disadvantage that isoscalar and isovector mesons composed by $u$ and $d$
quarks are degenerate); (v) the effective quark model by Vijande \textit{et al.} (set V)
\cite{Vijande05}, based upon the effective exchange of $\pi$, $\sigma$, $\eta$ and $K$ mesons
between constituent quarks; and (vi) the lattice QCD calculation by the HSC (set LQCD)
\cite{Dudek11}. Lattice QCD calculation does not include strange mesons as the previous models, but
it includes exotics such as the isoscalar $J^{PC}=2^{+-}$ states.

\begin{table}
\begin{center}
\caption[]{$p$-values of the K-S test for the experimental data from RPP and 
the six theoretical spectra. The null hypotheses are
that the NNSD coincides with the distorted Wigner surmise ($ p_{\sst DW}$)
and the distorted Poisson distribution ($ p_{\sst DP}$).} \vspace*{0.3cm}

\begin{tabular}{c||c|c|c|c|c}
\hline
     &       &      &        &              &               \\
Set  &    $d$  &  $n$  &  $d-n$   & $ p_{\sst DW}$ & $ p_{\sst DP}$ \\  \hline
     &       &      &        &              &               \\
RPP (data)  & $126$ & $23$ & $103$  &    $0.47$    &    $0.13$     \\ 
GI   &  $68$ & $17$ &  $51$  &    $0.84$    &    $0.41$     \\
K1   & $162$ & $38$ & $124$  &   $0.038$    &    $0.55$     \\
K2   & $162$ & $38$ & $124$  &   $0.005$    &    $0.43$     \\
E    & $190$ & $34$ & $156$  &   $0.083$    &    $0.21$     \\
V    &  $94$ & $18$ &  $76$  &    $0.51$    &    $0.56$     \\
LQCD &  $60$ & $15$ &  $45$  &   $0.033$    &    $0.44$     \\ \hline
\end{tabular}
\label{tableKS}
\end{center}
\end{table}

Table \ref{tableKS} displays relevant information on the six theoretical spectra, like their
dimension $d$, the number $n$ of pure sequences included in the analysis and the total number of
spacings, which is equal to $d-n$. It also provides the $p$-values obtained by applying the K-S test
to their NNSDs, taking as null hypotheses that the NNSD coincides either with $P_{DW}(s)$ or with
$P_{DP}(s)$. The first relevant outcome is that, according to the K-S test, the NNSDs
of sets K1, K2, E and LQCD are incompatible with the Wigner correlations and closer to the Poisson
statistics. Thus, the dynamics predicted by these models is essentially regular, while the
statistical properties of the experimental light meson spectrum show that the dynamical regime
should be chaotic.  This fact resembles the results obtained for baryons: while the fluctuations
of the experimental baryon spectrum are well reproduced by Wigner predictions, the theoretical
calculations give rise to spectra with Poisson statistics \cite{Fernandez07}.

\begin{figure}
\begin{center}
\rotatebox{0}{\scalebox{0.55}[0.55]{\includegraphics{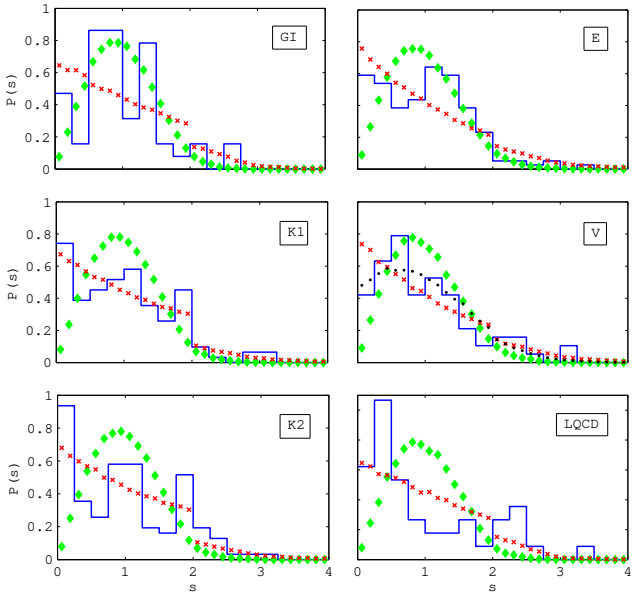}}}
\end{center}

\caption{(color online.) NNSDs for the theoretical spectra: i) Top left, set GI by Godfrey and Isgur
  \cite{Godfrey85}; ii) middle left, set K1 by Koll \textit{et al.} \cite{Koll00}; iii) bottom left,
  set K2 by Koll \textit{et al.} \cite{Koll00}; iv) top right, set E by Ebert \textit{et al.}
  \cite{Ebert09}; v) middle right, set V by Vijande \textit{et al.} \cite{Vijande05} and \textit{ad
    hoc} distorted Berry-Robnik with $f=0.63$; vi) bottom right, set LQCD by Dudek \textit{et al.}
  \cite{Dudek11}.  Distorted Wigner (D-Wigner) is represented with diamonds, distorted Poisson
  (D-Poisson) with crosses and distorted Berry-Robnik (D-Berry-Robnik) with dots.}
\label{NNSDteor}
\end{figure}

Figure \ref{NNSDteor} shows the NNSD of the six theoretical spectra. Sets K1 and E provide flat NNSDs with a cut at $s=2$.  The cut is expected as was explained in
section \ref{ssec:results}.  When the Poisson distribution is distorted it flattens due to the
small amount of levels, so actually the NNSDs that we find for sets K1 and E are the ones we expect
from a Poisson distribution, confirming that these sets have less correlations than the experimental
data as the K-S test suggested.  The comparison between models K1 and K2 by Koll \textit{et al.} is
particularly interesting because they only differ on the confinement interaction and show how
important that interaction can be for the spectral statistics, hinting that it should be revised to obtain a better agreement with the experiment.

The result for set LQCD is particularly interesting because lattice QCD is currently the only tool
available to compute low-energy observables employing QCD directly.  We find that the current
state-of-the-art calculation in \cite{Dudek11} does not describe properly the statistical properties
of the meson spectrum. Lattice QCD NNSD is relatively close to the $P_{DP}(s)$ as it is
  shown in figure \ref{NNSDteor}.  This is evident at zero spacing where $P_{LQCD}(s=0) \approx
  0.6$.  The value at zero spacing is critical to distinguish the Wigner and Poisson cases, being
  characteristic of a correlated spectrum and has an important impact in the K-S test.  Our results
  remain unaltered if the statistical errors of the lattice QCD calculation are taken into account.
Thus, the LQCD calculation should be considered a step forward in lattice calculations but still far
away from being a description of the data or their structure.  It is not something unexpected given
that the LQCD set has been obtained at a pion mass of 396 MeV, far away from its physical mass, and
it is reasonable to expect a drastic change in the statistical properties when calculations get the
pion mass closer to its actual value. However, the fact that the lattice QCD calculation has a lot
less correlations than the experimental data, being practically uncorrelated, demands, besides the
need of bringing the calculation closer to the physical pion mass, a careful examination of the
approximations employed. 

The results of the K-S test for sets GI and V are inconclusive because they
suggest that the models are compatible with both chaotic and integrable
dynamics. It is thus mandatory to take a close look
  to the NNSDs (figure \ref{NNSDteor}) before obtaining any conclusion.  Set GI NNSD has a strange
shape with peaks and dips, completely different from any of the usual distributions (Poisson, Wigner or Berry-Robnik), and therefore not close
at all to experiment (see figure \ref{NNSDexp2}). It only has some similarity
with some very particular integrable systems whose $P(s)$ is equal to a
sum of $\delta$ functions, constituting an exception to the rule of Poisson
distribution \cite{Makino03}. On the contrary, set V displays a smooth NNSD,
which can be very well fitted to a distorted Berry-Robnik distribution with
$f=0.63 \pm 0.19$ (also displayed in figure \ref{NNSDteor}). Then we can
conclude that the model by Vijande {\em et al.} gives a better account of the
dynamical regime of the light meson spectrum.

\section{Conclusions}\label{sec:conclusions}

In this letter we have studied the spectral fluctuations of the light meson
energy spectrum. These properties give
information about the dynamical regime of the physical system, i.e., whether
the system is chaotic, integrable or intermediate, and thus provide insight
on the underlying interactions. Our analysis
seems to be fairly robust against to the experimental errors, and unveils
that the dynamical regime of the light mesons is near to the fully chaotic
limit ---our results show that the spectral fluctuations of light mesons
resemble those of chaotic systems and indeed their NNSD can be well fitted to
a Berry-Robnik distribution with a 78\% of chaos.

On the other hand, the analysis of the spectra of the LQCD calculation and all
the quark models but one shows that they are incompatible with chaos.
This result is quite similar to that obtained for the baryons in Ref.
\cite{Fernandez07}. Only the model by Vijande {\em et al.} predicts an
intermediate spectrum with a 63\% of chaos. The other quark models and the
LQCD calculation~\cite{Koll00,Ebert09,Godfrey85,Dudek11} predict a regular
or nearly regular dynamics in clear contradiction with the experiment.

We would also like to point out that the experimental spectrum can be totally
chaotic, but the possible existence of missing levels, responsible for losing
correlations, would deflect the actual NNSD towards a distribution between
Wigner and Poisson. Ongoing experimental research will help to establish how
large is the effect of the missing states in the spectral-statistics
properties of the meson spectrum.

Further work is needed to study the origin of the failure of theoretical
models. In the case of the LQCD spectrum it is probably due to the fact that
the calculation is made at an unrealistic pion mass. For the quark models,
the ratio between the quark-antiquark interaction and the confinement
potential should be surveyed as well as the existence of states
that go beyond quark-antiquark content.

\section*{Acknowledgements}
The authors thank F.J. Llanes-Estrada, L. Scorzato and L.M. Fraile for their valuable
comments, J.J. Dudek for making available the HSC lattice QCD meson spectrum in \cite{Dudek11}, and
Professor F. Fern\'andez for providing an updated spectrum of the model in \cite{Vijande05}.  This
research has been conducted with support of the Spanish Ministry of Science and Innovation grants
FIS2009-11621-C02-01 and FIS2009-07277, and by CPAN, CSPD-2007-00042@Ingenio2010.  C.F.-R. is
supported by \textquotedblleft Juan de la Cierva\textquotedblright \: program of the Spanish Ministry of
Science and Innovation (Spain).  A.R. is supported by the Spanish program JAE-Doc (CSIC).

\end{document}